\documentclass[twocolumn,letter]{jpsj3} 

\usepackage{amsmath}

\title{Existence of Fine Structure inside Spin Gap in CeRu$_2$Al$_{10}$}

\author{
Hiroshi \textsc{Tanida}\thanks{E-mail address: tany@hiroshima-u.ac.jp}, 
Daiki \textsc{Tanaka}, Masafumi \textsc{Sera}, Chikako \textsc{Moriyoshi}$^1$, Yoshihiro \textsc{Kuroiwa}$^1$\\ Tomoaki \textsc{Takesaka}$^2$, Takashi \textsc{Nishioka}$^2$, Harukazu \textsc{Kato}$^2$ and Masahiro \textsc{Matsumura}$^2$
}

\inst{
Department of Quantum Matter, ADSM, Hiroshima University, Higashi-Hiroshima, Hiroshima 739-8530

$^1$Department of Physics, Hiroshima University, Higashi-Hiroshima, Hiroshima 739-8530

$^2$Graduate School of Integrated Arts and Science, Kochi University, Kochi 780-8520
}

\abst{

We investigate the magnetic field effect on the spin gap state in CeRu$_2$Al$_{10}$ by measuring the magnetization and electrical resistivity. 
We found that the magnetization curve for the magnetic field $H{\parallel}c$ shows a metamagnetic-like anomaly at $H^*{\sim}$4 T below $T_0$=27 K, but no anomaly for $H{\parallel}a$ and $H{\parallel}b$. 
A shoulder of the electrical resistivity at $T^s{\sim}$5 K for $I{\parallel}c$ is suppressed by applying a longitudinal magnetic field above 5 T. 
Many anomalies are also found in the magnetoresistance for $H{\parallel}c$ below ${\sim}$5 K. 
The obtained magnetic phase diagram consists of at least two or three phases below $T_0$. 
These results strongly indicate the existence of a fine structure at a low energy side in a spin gap state with the excitation energy of 8 meV recently observed in the inelastic neutron scattering experiments.

}

\kword{CeRu$_2$Al$_{10}$, spin gap, singlet ground state, magnetization, electrical resistivity}

\begin{document}
\maketitle

A ternary rare-earth compound CeRu$_2$Al$_{10}$, which crystallizes in the orthorhombic YbFe$_2$Al$_{10}$-type structure \cite{Verena,Tursina} has attracted much attention because of the mysterious long-range order (LRO) below $T_0$=27 K. 
First, the LRO below $T_0$=27 K was considered as the magnetic ordered phase. \cite{Strydom}
Contrary to the magnetic ordering,\cite{Strydom} Nishioka and Matsumura $et$ $al$. proposed the possibility of the nonmagnetic LRO below $T_0$ from the experimental results of the single crystal. 
They obtained the following results. 
(i) a large magnetic entropy of $S{\sim}0.7R$ln2 at $T_0$ accompanied with a large and sharp peak. \cite{Nishioka}
(ii) a decrease of the magnetic susceptibility below $T_0$ along all the crystal axes. \cite{Nishioka}
(iii) no internal magnetic field below $T_0$ probed by $^{27}$Al NQR. \cite{Matsumura}
(iv) a thermally activated behavior with activation energy of ${\sim}$100 K below $T_0$ observed in specific heat, magnetization ($M$), electrical resistivity ($\rho$), and nuclear spin relaxation rate of $^{27}$Al NQR. \cite{Nishioka, Matsumura}
(v) a positive pressure dependence of $T_0$ as $dT_0/dP$$>$0 below 2 GPa. \cite{Nishioka}
Nishioka $et$ $al$. proposed the charge density wave (CDW) formation \cite{Nishioka} and Matsumura $et$ $al$. proposed the structural phase transition. \cite{Matsumura}

In our previous paper, we proposed the singlet ground state from the above results and also the results of Ce$_x$La$_{1-x}$Ru$_2$Al$_{10}$. \cite{La-dope}
We reported that $T_0$ is suppressed by La-doping and also by applying the magnetic field along the magnetization easy axis, which clearly verify that the origin of the LRO is magnetic. \cite{La-dope}
Recently performed inelastic neutron scattering experiments \cite{INS} revealed the appearance of the spin gap with magnetic excitation energy of 8 meV below $T_0$. 
This strongly supports the LRO with the singlet ground state below $T_0$ proposed by us. \cite{La-dope} 
Thus, CeRu$_2$Al$_{10}$ is confirmed to be the first example of the LRO with singlet ground state in the $f$-electron compounds. 
However, the singlet ground state below $T_0$ is not so simple as is seen in the 3$d$ compounds. \cite{Hase,Hiroi}
Recently performed zero-field ${\mu}$SR experiments revealed the appearance of tiny internal field at the muon site below $T_0$. \cite{Kambe}
The temperature dependence of magnetic susceptibility shows a large Van Vleck contribution below $T_0$ whose magnitude depends on the applied field direction. \cite{Nishioka}
Subsequently, Hanzawa's investigation on the dimer of two Ce ions \cite{Hanzawa} whose crystalline electric field (CEF) ground doublet shows a large magnetic anisotropy indicates the possible existence of a large Van Vleck term in the singlet ground state and also an $H$-linear increase of $M$ with a large slope of $M$/$H$ which is different from the nonexistence of the Van Vleck term in the singlet ground state in the 3$d$ compounds. 
In our previous paper, \cite{transport} we reported the results of thermal conductivity, thermoelectric power, and Hall effect and the existence of the large anisotropy in these transport properties and pointed out that the system may have a two dimensional character in the $ac$-plane and we discussed the possibility of the singlet formation in the $ac$-plane.

In this letter, we investigated $M$ and $\rho$ of CeRu$_2$Al$_{10}$ single crystals in magnetic fields below $T_0$ in detail and found their characteristic anomalies for $H{\parallel}c$ in the LRO phase, which indicates not a simple singlet ground state in this compound.

CeRu$_2$Al$_{10}$ single crystal was prepared by the Al self-flux method. 
The obtained single crystals are typically a rectangular shape. 
On the surface of $b$-plane, an extra lump with a pyramid shape is frequently found. 
The grown-axis is different between the rectangle and the pyramid. 
In the following measurements, the samples were used after eliminating the extra part from the rectangle. 
$M$ was measured by a SQUID magnetometer down to 2 K up to 5 T. 
$\rho$ was measured by a conventional ac four-probe method down to 0.4 K up to 14.5 T.

Figure 1 shows the temperature ($T$) dependence of the magnetic susceptibility ($\chi$=$M/H$) of CeRu$_2$Al$_{10}$ at $H$=1 T along the three crystal axes. 
The easy and hard axes of $M$ are $a$- and $b$-axes, respectively. 
Note that $\chi_a$ ($H{\parallel}a$) and $\chi_c$ ($H{\parallel}c$) in ref. 4 (Fig.2 (a)) are correctly $\chi_c$ and $\chi_a$, respectively; see ref.7 in ref.4. 
Although the result of $\chi_a$ is almost the same as our previous results, \cite{Nishioka, La-dope} those of $\chi_b$ ($H{\parallel}b$) and $\chi_c$ are slightly different from those reported in our previous paper. \cite{Nishioka}
This is because the sample in our previous paper \cite{Nishioka} was $not~a~perfect~single~crystal$. 
In the $bc$-plane, $b$- and $c$-axes are mixed together. 
In the present study, we used the almost single crystals, which were used also in our previous studies of transport properties. \cite{transport}
Within confirmations of Laue photograph, no additional spots due to the mixture of these axes were observed.
As for $\chi_a^{-1}$ and $\chi_c^{-1}$, a Curie-Weiss behavior is observed at high temperatures. 
For $\chi_b^{-1}$, a broad maximum is seen at $T{\sim}$250 K, suggesting that the CEF excited states are situated above several hundred Kelvins. 
At low temperatures, a small upturn is observed in all of $\chi_a$, $\chi_b$, and $\chi_c$.

\begin{figure}[tb]
\begin{center}
\includegraphics[width=0.7\linewidth]{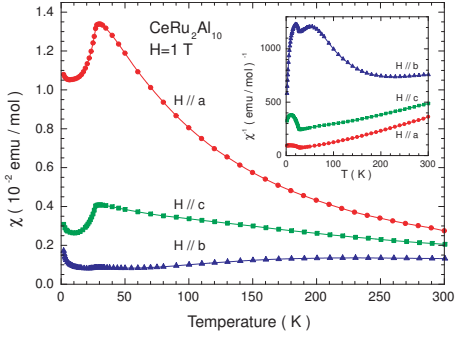}
\end{center}
\caption{
(Color online) The $T$ dependence of magnetic susceptibility of CeRu$_2$Al$_{10}$. 
In the inset, the inverse magnetic susceptibility is plotted as a function of the temperature. 
}
\label{f1}
\end{figure}

Figures 2(a), 2(b), and 2(c) show the $M$-$H$ curves of CeRu$_2$Al$_{10}$ for $H{\parallel}a$, $H{\parallel}b$, and $H{\parallel}c$, respectively. 
$M$ for $H{\parallel}a$ ($M_a$) shows almost an $H$-linear increase independent of temperature. 
$M$ for $H{\parallel}b$ ($M_b$) at low temperatures shows a convex $H$ dependence, while those at high temperatures show an $H$-linear increase. 
These results of $M$-$H$ curves for $H{\parallel}b$ are explained by the existence of a small amount of magnetic impurities.
As shown in Fig. 2(c), $M$ for $H{\parallel}c$ ($M_c$) at $T$=30 K shows a simple $H$-linear increase as expected in a paramagnetic region. 
However, below $T_0$ down to 2 K, an anomaly is seen at $H{\sim}$4 T. 
The anomaly of $M$-$H$ curve looks like a metamagnetic like or a spin-flop like transition at first glance. 
We define the magnetic field showing the anomaly as $H^*$, in which $H^*$ is defined as a field where $dM/dH$-$H$ curve shows a maximum as shown in the inset of Fig. 2(c). 
The anomaly around $H^*$ is the clearest at $T{\sim}$10 K as is seen in the $dM/dH$-$H$ curve. 
Below $H\sim$2 T, $M_c$ shows an almost $H$-linear increase, which indicates that the effect of the magnetic impurity is small down to 2 K. 
While the $M$-$H$ curve for $H{\parallel}c$ shows the anomaly below $T_0$, those for $H{\parallel}a$ and $H{\parallel}b$ show no anomaly up to 14.5 T and 5 T below $T_0$, respectively. 
Thus, the anomalous $M$-$H$ curve at low magnetic fields below $T_0$ is a specific property for $H{\parallel}c$.

\begin{figure}[tb]
\begin{center}
\includegraphics[width=0.9\linewidth]{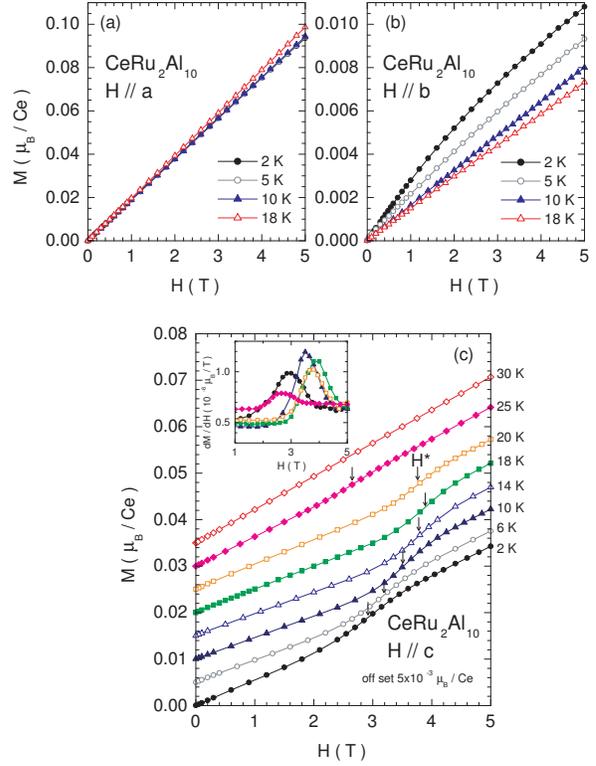}
\end{center}
\caption{
(Color online) The $M$-$H$ curves of CeRu$_2$Al$_{10}$ for (a) $H{\parallel}a$, (b) $H{\parallel}b$, and (c) $H{\parallel}c$, respectively. 
$H^*$ is illustrated by arrows. 
In the inset of (c), $dM/dH$ is shown for 2, 10, 18, 20, and 25 K.

}
\label{f1}
\end{figure}

Figure 3(a) shows the $T$ dependence of $M/H$ of CeRu$_2$Al$_{10}$ measured at $H$=1 and 5 T along the $a$-axis. 
The upturn of $M_a/H$ observed at $H$=1 T is completely suppressed by applying the magnetic field of $H$=5 T. 
Similar tendency is also observed in $M_b/H$ as shown in Fig. 3(b). 
The upturns of $M_a/H$ and $M_b/H$ at low temperatures are ascribed to the existence of a small amount of magnetic impurities. 
Figure 3(c) shows the $T$ dependence of $M/H$ of CeRu$_2$Al$_{10}$ for $H{\parallel}c$. 
At $H$=1 T, $M_c/H$ shows a steep decrease below $T_0$, and after exhibiting a minimum at $T{\sim}$10 K, increases down to 2 K. 
While the temperature of the minimum increases up to $T{\sim}$13 K with increasing magnetic field up to $H$=3 T, the $M/H$-$T$ curves coincide with that at $H$=1 T above $T{\sim}$13 K. 
Such an increase of $M_c/H$ at low temperatures would not be ascribed to the magnetic impurity. 
In a usual increase originating from the magnetic impurity, the lower the magnetic field, the larger the upturn from the Curie term as seen for $H{\parallel}a$ and $H{\parallel}b$. 
However, in the present case for $H{\parallel}c$, the opposite tendency is observed. 
Thus, the $T$ dependence of $M/H$ for $H{\parallel}c$ at low temperatures is the intrinsic property in this compound. 
At $H$=3.4 and 4 T, the shoulders are seen at $T^*{\sim}$24 K and 22 K, respectively. 
These corresponds to the anomaly at $H^*$ in the $M$-$H$ curve at these two temperatures. 
At $H$=5 T, a large suppression as is observed below 3.4 T is not seen in the $T$ dependence of $M_c/H$ below $T_0$ and shows a broad and small minimum at $T{\sim}$15 K. 
This is associated with the $H$-linear $M$-$H$ curve above ${\sim}H^*$ below $T_0$.

\begin{figure}[tb]
\begin{center}
\includegraphics[width=0.7\linewidth]{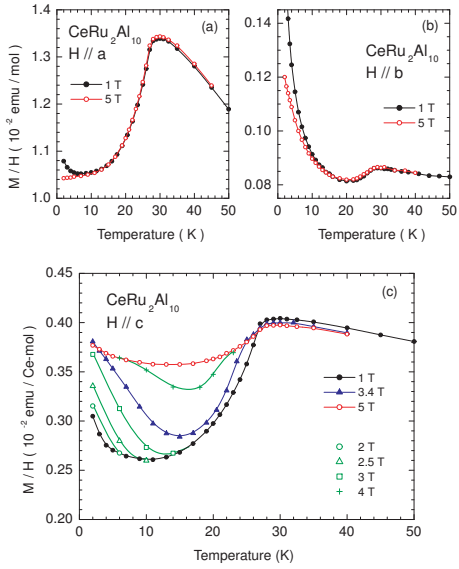}
\end{center}
\caption{
(Color online) The $T$ dependence of $M/H$ of CeRu$_2$Al$_{10}$ for (a) $H{\parallel}a$, (b) $H{\parallel}b$, and (c) $H{\parallel}c$, respectively. 
As for $H{\parallel}c$, data obtained from $M$-$H$ curves are also plotted for $H$=2, 2.5, 3, and 4 T. 

}
\label{f1}
\end{figure}

Figure 4 shows the $T$ dependence of $\rho$ of CeRu$_2$Al$_{10}$ under the longitudinal magnetic field along the $c$-axis. 
The magnetic field effect at low temperatures is much smaller than that under the transverse magnetic field reported in our previous paper. \cite{La-dope}
Above $\sim$20 K, a small negative magnetoresistance is observed. 
This originates from the suppression of the conduction electron scattering by the localized moments. 
The smaller magnitude of the negative magnetoresistance than that for $H{\parallel}a$ \cite{La-dope} originates from the smaller magnetization for $H{\parallel}c$. 
Although at $H$=0, a small shoulder is seen at $T^s{\sim}$5 K, it is not seen in magnetic fields above $H{\sim}$5 T. 
This is more clearly seen in the inset of Fig. 4 where the $T$ dependence of $\rho/\rho_{10 \rm K}$ is shown. 
$\rho_{10 \rm K}$ means $\rho$ at $T$=10 K under each magnetic field. 
This suggests that a small shoulder of $\rho$ at $T^s{\sim}$5 K comes from the magnetic origin.

\begin{figure}[tb]
\begin{center}
\includegraphics[width=0.7\linewidth]{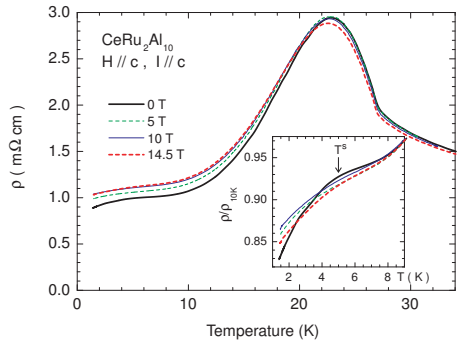}
\end{center}
\caption{
(Color online) The $T$ dependence of electrical resistivity of CeRu$_2$Al$_{10}$ for $H$ and $I{\parallel}c$.
The inset shows the low temperature expansion in the form of ${\rho}/{\rho}_{10}$$_{\rm K}$. 
$T^s$ is illustrated by an arrow. 
}
\label{f1}
\end{figure}

Figure 5 shows the longitudinal magnetoresistance of CeRu$_2$Al$_{10}$ along the $c$-axis. 
The inset shows those at high temperatures in the form of $\rho/\rho_0$. 
Here, $\rho_0$ is $\rho$ in zero magnetic field at each temperature. 
Between 10 K and 25 K, one anomaly is seen around $H{\sim}$4 T. 
This corresponds to $H^*$ observed in the $M$-$H$ curve along the $c$-axis shown in Fig. 2(c). 
At $T$=25 K, a negative magnetoresistance is observed and an anomaly is seen at $H{\sim}$2.5 T. 
Below $\sim$20 K, a sign of magnetoresistance changes from negative to positive. 
$\rho$ at $T$=18 K, 15 K, and 10 K show the similar $H$ dependence: a concave $H$ dependence below $H^*$ and a convex one above $H^*$. 
This indicates that the conduction electron scattering is small below $H^*$ and large above $H^*$, and suggests that the nature of singlet ground state below $H^*$ is different from that above $H^*$. 
Below ${\sim}$7 K, many anomalies appear. 
The anomaly at $H{\sim}$4 T at 10$\sim$18 K seems to be separated into two anomalies below $\sim$7 K, and smeared out below $\sim$2 K. 
Below $T{\sim}$5 K, many anomalies are seen at high magnetic fields, and the lower the temperature the more pronounced the anomalies, which rather seems a oscillatory behavior. 
Below $T{\sim}$3 K, a saturated behavior is seen above $H{\sim}$10 T. 
Under the transverse magnetic field, such a complicated behavior was not found. \cite{La-dope}
Thus, the metamagnetic anomalies at $H^*$ are specific for $H{\parallel}c$.

We note the anomalies observed in the magnetoresistance at low temperatures. 
As the possible origin of these anomalies at low temperatures, both extrinsic and intrinsic effects are considered. 
For the oscillatory behavior at low temperatures, the Shubnikov-de Haas (SdH) effect of Al which was used as a flux. 
However, the magnetoresistances at $T$=0.44K and 0.8K show no anomaly around 100 Oe which is expected from the superconductivity of free Al ($T_c$=1.2K). 
Although the possibility of SdH effect of Al cannot be ruled out, its possibility may be very small because the magnitude of the anomaly observed in this experiments is expected to be much larger than the residual resistivity of Al used as a flux. 
As an intrinsic effect, the SdH effect of CeRu$_2$Al$_{10}$ itself is also possible, however, that is also unlikely because the no periodicity is determined within the maximum magnetic fields. 
Thus, we consider that these anomalies are intrinsic in CeRu$_2$Al$_{10}$, and come from the fine structure in a spin gap.

\begin{figure}[tb]
\begin{center}
\includegraphics[width=0.65\linewidth]{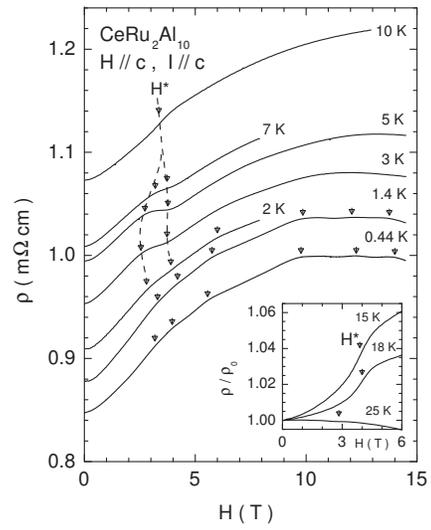}
\end{center}
\caption{
The magnetoresistance of CeRu$_2$Al$_{10}$, where $H{\parallel}c$ and $I{\parallel}c$. 
In the inset, those measured at high temperatures are shown in the form of ${\rho}/{\rho}_0$, where ${\rho_0}$ is ${\rho}$ at $H$=0. 
}
\label{f1}
\end{figure}

Figure 6 shows the magnetic phase diagram of CeRu$_2$Al$_{10}$ for $H{\parallel}c$. 
We define the ground state and the high-$T$ phase above $H^*$ as phase A and B. 
Below ${\sim}$10 K, $H^*$ seems to be separated into two different critical fields. 
As shown in Fig. 6, a broad boundary exists at low magnetic fields around $T^s{\sim}$5 K where $\rho$ shows a shoulder. 
Below $T{\sim}$3 K, many anomalies appear at high magnetic fields. 
Thus, the magnetic phase diagram of CeRu$_2$Al$_{10}$ for $H{\parallel}c$ is very complicated, which indicates the existence of the internal structures in the singlet ground state.

\begin{figure}[tb]
\begin{center}
\includegraphics[width=0.7\linewidth]{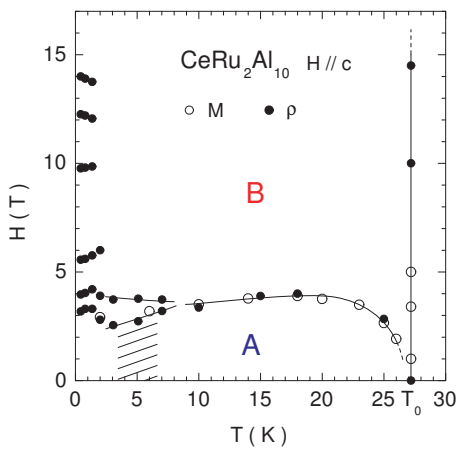}
\end{center}
\caption{
(Color online) The $H$-$T$ phase diagram of CeRu$_2$Al$_{10}$ for $H{\parallel}c$. 

}
\label{f1}
\end{figure}

Here, we discuss the anomalous $T$ dependence of $\rho$ around $T^s$ at $H$=0; 
a shoulder for $I{\parallel}b$ and $I{\parallel}c$ and a broad maximum for $I{\parallel}a$ around $T^s$. \cite{Nishioka}
The thermoelectric power $S$ at $H$=0 also exhibits the anomaly around $T^s$ along all the crystal axes; a broad maximum in $S_a$, $S_b$, and $S_c$, respectively. \cite{transport}
A shoulder of $\rho$ around $T^s$ for $I{\parallel}c$ disappears in magnetic fields above 5 T for $H{\parallel}c$. 
The suppression of a maximum around $T^s$ in $S$ of the polycrstal by magnetic field was also reported. \cite{Strydom2}
Thus, the origin of the anomalies of $\rho$ and $S$ around $T^s$ is considered to be magnetic and its energy scale is expected to be ${\sim}T^s$. 
This energy scale of ${\sim}T^s$ seems to be comparable with that of $H^*$. 
Different from the usual metamagnetic transition where the magnitude of the anomaly increases with decreasing temperature, in the present case, it is the largest at around 10 K, and $H^*$ is the highest at around 18 K and both decrease with decreasing temperature. 
The slope of $M/H$ at low magnetic fields is the smallest around the intermediate temperature of $\sim$10 K and increases with decreasing temperature. 
This is seen also in the $T$ dependence of $M$ where a Curie like upturn is enhanced with increasing magnetic field up to ${\sim}$3 T. 
At a glance, the magnitude of a spin gap seems to be rather small if we estimate it from the $M$-$T$ curve at $H$=5 T, which is different from ${\sim}$100 K estimated at $H$=1 T. 
However, the $T$ dependence of $\rho$ below $T_0$ is nearly the same up to $H$=14.5 T. 
This suggests that the magnitude of a spin gap is not changed below and above $H^*$. 
By considering that it is confirmed that the spin gap with the magnetic excitation energy of 8 meV is formed below $T_0$ observed in the inelastic neutron scattering, \cite{INS} the present results indicate the existence of the internal magnetic fine structure at a low energy side inside a spin gap. 
A small internal field observed in zero-field $\mu$SR experiment \cite{Kambe} is possible to be associated with this fine structure. 
This ingap structure has a nature that is easily destroyed for $H{\parallel}c$. 
As the easy magnetization axis is along the $a$-axis, a magnetic moment along this direction seems to be difficult to be induced by destroying a singlet ground state. 
The $c$-axis is a hard magnetization axis in the $ac$-plane. 
This is possible to be associated with the induced magnetization above $H^*$ for $H{\parallel}c$ as if the magnetization is easy to be induced in a $\chi_{\perp}$ configuration in a usual antiferromagnet. 
At present stage, the details of the internal fine structure inside a spin gap are not known. 
Such an unusual spin gap state different from that in 3$d$ compounds may originate from the orbital degrees of freedom in the singlet ground state in $f$-electron system.

To conclude, we studied the magnetization and magnetoresistance of CeRu$_2$Al$_{10}$. 
We found the anomaly in the magnetization and magnetoresistance at $H^*{\sim}$4 T for $H{\parallel}c$ below $T_0$. 
The anomaly is not seen in the $M$-$H$ curve for $H{\parallel}a$ and $H{\parallel}b$ at least up to 14.5 T and 5 T, respectively. 
A shoulder of $\rho$ at $T^s{\sim}$5 K for $H{\parallel}c$ is suppressed by applying a magnetic field above 5 T. 
The magnetoresistance for $H{\parallel}c$ shows many anomalies below $\sim$5 K. 
From these results, we obtained the unusual magnetic phase diagram which contains at least two or three phases below $T_0$. 
These may be considered to originate from the existence of a fine structure at a lower energy side in a spin gap. 
Thus, the present study indicates the unusual ground state within a spin gap below $T_0$ in CeRu$_2$Al$_{10}$.

\end{document}